\documentstyle[aps,prl,epsfig]{revtex}

\begin{document}

\twocolumn[\hsize\textwidth\columnwidth\hsize\csname @twocolumnfalse\endcsname
\widetext 

\title{Dilation-induced phases of gases absorbed within a bundle of carbon nanotubes}
\author{M. Mercedes Calbi$^{a}$, Flavio Toigo$^{a,b}$ and Milton W. Cole$^{a}$}
\address{$^{a}$Department of Physics, Pennsylvania State University, University Park, PA 16802 USA}
\address{$^{b}$INFM and Dipartimento di Fisica ''G. Galilei'', via Marzolo 8,I-35131, Padova, Italy}
\date{\today}
\maketitle

\begin{abstract}
A study is presented of the effects of gas (especially H$_{2}$) absorption
within the interstitial channels of a bundle of carbon nanotubes. The ground
state of the system is determined by minimizing the total energy, which
includes the molecules' interaction with the tubes, the inter-tube
interaction, and the molecules' mutual interaction (which is screened by the
tubes). The consequences of swelling include a significant increase in the
gas uptake and a 3 per cent increase in the tubes' breathing mode frequency.

\bgroup\draft
\egroup
\end{abstract}
] \narrowtext 
Considerable attention has focused in recent years on the absorption of H$%
_{2}$ (and other gases) in various forms of carbon, especially high surface
area materials, because of the potential such materials present for
efficient storage, isotope separation and other applications. Numerous
techniques of interfacial science are being applied to study the remarkable
properties of this system. In the case of carbon nanotubes (NT), the unusual
geometry presents the possibility of novel phase transitions. Theoretical
work has explored a variety of transitions and proposed ways to observe them
experimentally \cite{cole,gordillo,krotscheck}.

In this paper we propose such an unusual transition for hydrogen and other
gases, associated with the dilation of a lattice of nanotubes. The most
striking feature of a single nanotube is its quasi-one dimensional (1D)
character, which makes contact with a growing body of 1D theory \cite{tak}.
This geometry stimulated Gordillo, Boronat and Casulleras \cite{gordillo} to
study a system of H$_{2}$ molecules in 1D and quasi-1D (in which case the
molecules' small amplitude motion perpendicular to the tube's axial (z)
direction was taken into account). They found density-dependent transitions
at temperature $T=0$ to a liquid phase and then to a novel 1D solid phase.
These results are qualitatively similar to those found by Boninsegni and
coworkers \cite{cole,boninsegni} for $^{4}$He in both 1D and within an
ordered lattice (bundle) of parallel NTs. If we consider such a bundle, then
the interactions between molecules in adjacent interstitial channels (IC)
allows a fully 3D transition to occur. However, the large spacing ($\sim 10$ 
\AA ) between adjacent channels means that the inter-IC interaction
is weak and the predicted ordering temperature T$_{c}$ is somewhat lower
than the energy scale set by the well depth $\varepsilon $ of the pair
potential \cite{cole}. The transition is one to a very anisotropic liquid
(in the case when the molecules are assumed to move freely along the
channels) or crystal, depending on the system \cite{carraro}.

Two new aspects of the absorption problem alter this situation both
qualitatively and quantitatively. One is that a study of three-body
interactions found\cite{screen} that an effect of the NTs' dynamic
polarization is to significantly screen the interaction between molecules,
with drastic consequences \cite{kostov}. If one includes only intra-IC
interactions, the 1D H$_{2}$ ground state in this environment is a very low
density gas instead of the high density, strongly self-bound liquid
mentioned above. However, inclusion of the inter-IC interactions does yield
a weakly bound condensed state \cite{kostov}. The other new ingredient in
this problem is some suggestion and tentative evidence \cite{bienfait,migone}%
, that adsorbed gases cause the bundle of NTs to swell. In this paper, we
consider the problem of such a dilation of the lattice of NTs. We report
four predictions associated with this dilation: a greatly increased binding
of the hydrogen, a significantly higher critical temperature for the
condensed state, a measurably larger lattice constant of the NT array, and a
higher breathing mode frequency for the tubes. Several other consequences
merit future study. Especially interesting is the effect of this dilation on
the electrical transport properties of the NTs, which have been found to be
significantly altered by adsorption \cite{trans1,trans2}. While we focus on
the case of H$_{2}$, we have found and summarize below qualitatively similar
conclusions for the adsorption of He, Ne,Ar, and CH$_{4}$. The unifying
concept is that lattice dilation permits small molecules to increase their
IC binding energy significantly without a substantial increase in the
inter-tube interaction energy.

The logic of our calculation is the following. Consider an infinite array of
\ NT's, each of which has radius $R$ and a large length $L$, and whose
parallel axes intersect an orthogonal plane to form a triangular lattice of
spacing $d_{0}=17$ \AA\ at $T=0$; we are ultimately interested in the
thermodynamic limit of infinite $L$. We assume that between the tubes,
within the IC's, there exists a 1D density $\rho $ of H$_{2}$ molecules, so
that on average there are $N=\rho L$ molecules per channel. We then minimize
the total energy of the system by allowing the lattice to dilate to a
separation $d$. Such a resulting situation turns out to be stable only if $%
\rho $ exceeds a threshold density, $\rho _{c}$. This threshold density and
attendant dilated lattice represent the ground state of the system. The
finding that any density lower than $\rho _{c}$ is impossible is analogous
to the familiar absence of a stable, uniform low density regime for
conventional systems in free space at $T=0$, i.e., the ground state of all
3D systems is a condensed liquid or solid. As in the latter case, at finite $%
T$ a low density gas phase does exist within the NT bundle, as does a regime
of coexistence between this dilute gas and the dense liquid.

The calculations themselves are straightforward, thanks to several
simplifying assumptions which should not greatly affect the major
conclusions; these include the neglect of any $z$ dependence in the
potential experienced by the H$_{2}$ molecules. The total energy of the
system, per unit length, per IC, indicated by $\varepsilon ,$ is written in
terms of the single molecule's energy per particle $(\epsilon _{1})$, of the
H$_{2}$-H$_{2}$ interaction energy per particle ($\epsilon _{int}$), and of
the elastic energy per unit length representing the interaction between
pairs of adjacent NTs as: 
\begin{equation}
\varepsilon \equiv E/L=\rho \lbrack \epsilon _{1}(d)+\epsilon _{int}(\rho
,d)]+\frac{1}{2}k(d-d_{0})^{2}  \label{eq_en}
\end{equation}

\ The coefficient $k=1740$ K \AA $^{-3}$ is derived from the semiempirical
NT interaction constant of Mizel et al \cite{mizel}. This value is
consistent with the experimental value of the interlayer force
constant and interaction energy of graphite.
The function $\epsilon
_{int}(\rho ,d)$ is our variational upper bound to the ground state energy
per particle of fully interacting H$_{2}$, computed with the screened
interaction. This quantity is the sum of the ''exact'' ground state energy
of 1D H$_{2}$ (computed by Boronat and Gordillo\cite{boronat}, using the
diffusion Monte Carlo method) and the small interchannel interaction,
computed in the Hartree approximation. Finally, $\epsilon _{1}(d)$ is the
ground state energy per H$_{2}$ molecule subjected to the potential energy
within an IC of a bundle of tubes which are spaced a distance $d$ apart. We
deduce this potential energy from the model of Stan and coworkers\cite{stan}
and solve the Schr\"{o}dinger equation using the diffusion method in order
to obtain $\epsilon _{1}(d)$.

\begin{figure}[tbh]
\centerline{\psfig{file=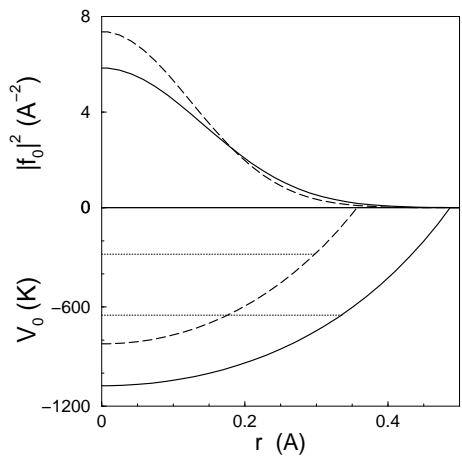, width=6cm }}
\caption{ Azimuthally averaged potential energy and ground state eigenvalue
(bottom panel) and probability density (top panel) of H$_{2}$ as a function
of perpendicular distance from the center of the interstitial channel.
Dashed curves correspond to the undilated NT lattice and full curves
correspond to the lattice dilated by 1 per cent (the predicted ground state
of the system). }
\label{fig1}
\end{figure}

\noindent

\begin{figure}[tbh]
\centerline{\psfig{file=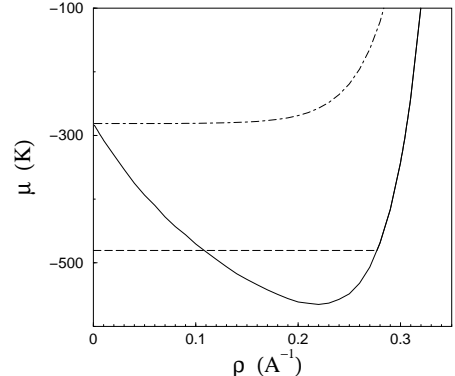, width=5cm, angle=-90}}
\caption{
Chemical potential as a function of the 1D density $\rho $ of H$_{2}$ (full curve). The horizontal line exhibits the equilibrium (two-phase) behavior obtained from Maxwell construction which determines the uptake threshold density $\rho_{c}$. Dash-dot curve is hypothetical result in the absence of dilation. 
}
\label{fig2}
\end{figure} 

Figure \ref{fig1} shows the potential energy and wave function for the
cases $d=d_{0}=17$ \AA\ and a slightly larger value, $d=17.18$ \AA . A key
difference between these two cases is that the ground state of the dilated
lattice has a much lower value of $\epsilon _{1}(d)$. 
The reasons for this
decrease are the lower potential energy and zero-point energy (ZPE)
associated with the increased distance of the H$_{2}$ molecules from their
neighboring NT's.
This decrease in $\epsilon _{1}$ (equivalently, the
quantum pressure, which in H$_{2}$ is near 100 atm at threshold) is what
drives the NT's apart \cite{pressure}. Figure \ref{fig2} shows the
chemical potential ($\mu =\frac{\partial E}{\partial N}=\frac{\partial
\varepsilon }{\partial \rho }$) of the hydrogen as a function of $\rho $,
taking into account the lattice dilation.

\begin{figure}[tbh]
\centerline{\psfig{file=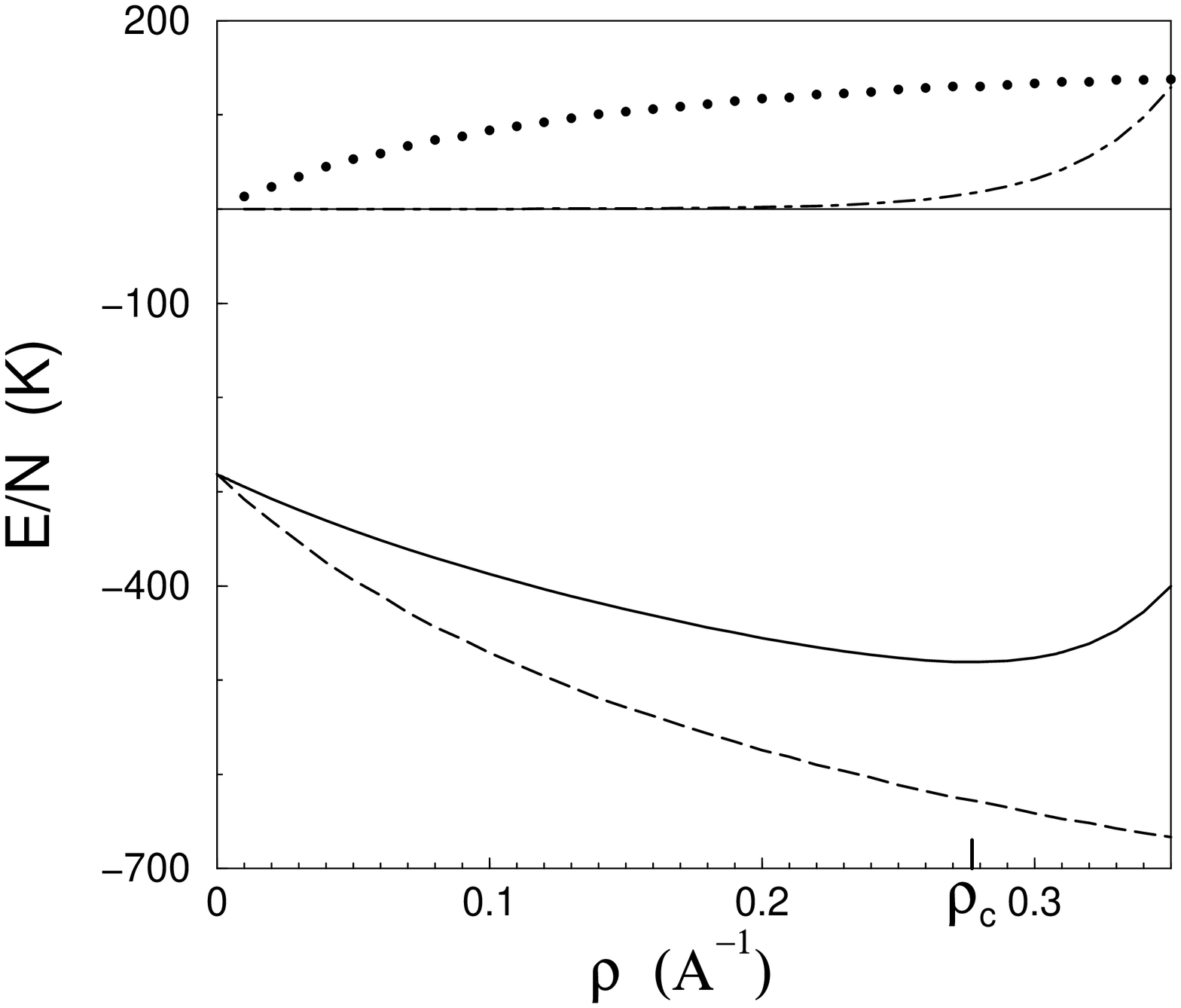, width=5cm, angle=-90}}
\caption{
Energies as a function of the 1D density $\protect\rho $ of H$_{2}$. Filled circle is the NT-NT elastic interaction energy, dash-dot curve is the  H$_{2}$-H$_{2}$ interaction energy  per molecule and dashed curve  represents the H$_{2}$ molecule's eigenvalue in the dilated lattice. Full curve is the sum of the above contributions. Notice that its minimum occurs at $\rho_c$. 
}
\protect\label{fig3}
\end{figure}

Also shown is the analogous result in the case of an undilated lattice. In
the latter case, the curve exhibits an extremely shallow minimum near $\rho
=0.1$ \AA $^{-1}$, which is not discernible in the figure; the binding
energy in that case is $\sim 0.1$K \cite{kostov}. In the present case
of dilation, the chemical potential is seen to decrease rapidly with
increasing $\rho $ because the expanding lattice becomes progressively more
attractive, as seen in Fig. \ref{fig1}; the eigenvalue of this potential
has a minimum at lattice constant $17.55$ \AA , corresponding to a very large density. However, the mutual H$_{2}$ repulsion at such high
density is huge. The ground state energy of the system is obtained by
minimizing the grand potential, $\omega =\varepsilon -\mu \rho $. This is
equivalent to a Maxwell construction in the chemical potential-density
plane, as seen in the figure. It corresponds equivalently to setting the 1D
pressure of the H$_{2}$ equal to the stress of the tubes, associated
with the change of their total interaction energy with respect to density.

\begin{figure}[tbh]
\centerline{\psfig{file=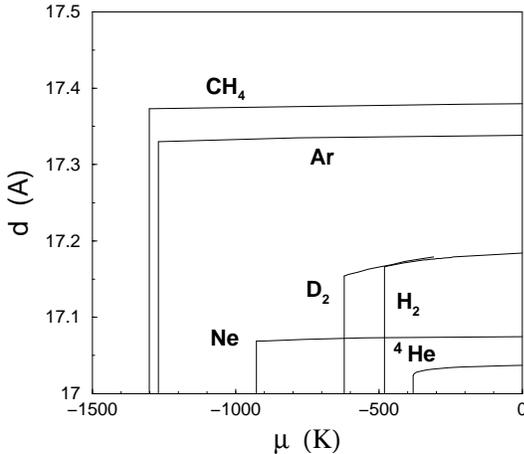, width=6cm,angle=-90}}
\caption{ 
Lattice constant of the NT array as a function of 1D density for $%
^{4}$He, H$_{2}$ and D$_{2}$. The discontinuities occur at the respective
T=0 thresholds for gas uptake.
 } 
\label{fig4}
\end{figure}

 The resulting values of the ground state density, lattice constant
and chemical potential, $\rho _{c}$, $d_{c}$ and $\mu _{c}\ \ $, are
reported in Table I.

\begin{table}[tbh]
\caption{ Density, NT lattice constant, and chemical potential at threshold for gas uptake.}

\begin{tabular}{|c|c|c|c|c|c|c|}
& $^{4}$He & Ne & H$_{2}$ & D$_{2}$ & Ar & CH$_{4}$ \\ \hline
$\rho _{c}$ (\AA $^{-1}$) & 0.215 & 0.327 & 0.277 & 0.275 & 0.273 & 0.270 \\ 
\hline
$d_c$ (\AA $^{{}}$) & 17.024 & 17.069 & 17.166 & 17.154 & 17.330 & 17.373 \\ 
\hline
$\mu_c$ $($K$)$ & -380.7 & -927.5 & -480.7 & -621.4 & -1270 & -1290 \\ 
\end{tabular}
\end{table}
\noindent

Note that the chemical potential at this threshold is about $200$ K lower than the
value $-281.5$ K found in the absence of dilation. This is a measure of the
greater incentive to absorb within this environment than within an undilated
NT bundle. Such a dramatic result should be testable by thermodynamic
measurements, either adsorption isotherms or direct measurements of the heat
of adsorption. We note that the computed value of $\mu _{c}$ is higher than
the corresponding threshold, near $\sim -600$K, for absorption within
the tubes (which is relevant only if the tubes are open) or in the grooves
formed on the external surface of the bundle. The difference is a
consequence of the much lower zero-point energy in these sites than in the
IC's. Figure \ref{fig3} shows 
the various contributions to the energy per particle as a function of $\rho$.

Figure \ref{fig4} presents our predictions for the dependence of $d$
on $\mu $ for various gases. The isotope
dependence arises because the smaller mass of H$_{2}$ implies a larger ZPE.
Hence, in the D$_{2}$ case the lattice is dilated less and the threshold
values of $\rho $ and the chemical potential are both lower than in the H$%
_{2}$ case \cite{sieve}. Note that the swelling is predicted to be $\sim 1\%$ change in the value of $d$ at the threshold for gas uptake.
This is measurable by X ray diffraction from the NT lattice 
\cite{lucas}. Table I reports results of the effects of dilation on various
gases. Two of these, He and Ne, fit nearly perfectly in the undilated
lattice; hence the dilation $d$ is less than $0.5\%$ and the increase in
binding is small ($\sim 3\%$ for Ne). For H$_{2}$ , Ar and CH$_{4}$ , the
energetic consequence of the dilation is significant. One factor to bear in
mind is the sensitivity of these results to the potential parameters. As examples, in the case of both H$_{2}$ and CH$_4$, a 2.5 percent in the gas-carbon  length parameter ( $\sigma_{GC}$) results in about a 25 percent increase in the magnitude of $\mu_c$.

The nanotubes should exhibit other consequences of the H$_{2}$ confinement.
One, possibly surprising, finding is a significant shift in the breathing
mode frequency of the tubes. To evaluate this, we consider the degree of
freedom corresponding to the variation in radius of the tubes, $R$.
''Breathing'' corresponds to a uniform radial expansion and contraction of a
tube. We evaluate the shift by adding an extra term $\delta \varepsilon (R)=%
\frac{1}{2}\gamma (R-R_{0})^{2}$ to the previous expression for the total
energy per unit length, Eq. \ref{eq_en}. Here $R_{0}$ is the equilibrium
radius in the absence of the H$_{2}$ and $\gamma =2.5 \cdot 10{^{5}}$ K \AA $^{-3}$ is the force constant derived from the breathing mode frequency of the
unperturbed lattice \cite{breathing}. Such a change in $R$ \ yields a change
in potential energy experienced by the H$_{2}$ molecules, which we compute
from the formulation of Stan et al.\cite{stan}. As expected by analogy with
the dilation problem, there ensues a small decrease in the tube radius (of
order $0.01$\AA ). Much more dramatic is the effect on the breathing
frequency, which can be evaluated by computing the second derivative of the
total energy with respect to $R$. Our calculations yield a $3\%$ change in
the breathing frequency at the threshold ($\rho =\rho _{c}$) for H$_{2}$
uptake at $T=0$. This should be easily measured spectroscopically \cite
{breathing}. The magnitude of the shift is a manifestation of the
sensitivity of the H$_{2}$ potential to the positions of the nearby carbon
atoms; it is comparable to the shift of this frequency due to tube-tube
interactions. 

We may also make some predictions concerning the finite temperature phase
diagram of this unusual system. Below a critical temperature T$_{c}$
there occurs a region of two-phase coexistence, between a low density
quasi-1D ''vapor'' phase, in a nearly unswollen NT environment, and a high
density phase in swollen ICs. At very low $T$, this latter phase is 
essentially
the ground state fluid, with a negligible thermal excitation so its chemical potential satisfies $\mu(\rho_c,T) \simeq \mu(\rho_c,0)$. The vapor
density at coexistence may be determined by equating the chemical
potentials, a procedure which is particularly simple at very low $T$. Since
this vapor phase is dilute, its chemical potential is well approximated
by the 1D ideal gas relation $\mu =k_{B}T\ln \rho \lambda +\epsilon _{1}(d)$
where $\lambda$ is the de Broglie wavelength of the molecules.
By equating these we determine the curve: $\rho _{v}\lambda
=e^{\beta (\mu _{l}(\rho _{c}(0)-\epsilon _{1}(d_{0}))}$  describing coexistence of the low density phase and the high density phase.
At temperatures above a critical value $T_{c}$, $\mu (\rho )$
will no longer exhibit a minimum and only one phase will be possible.

Lacking a better theory, we estimate $T_{c}$ by mapping our system to an
Ising system with weak long range forces with hamiltonian $H=-\frac{1}{2}%
\frac{J}{N}\sum_{i,j}\sigma _{i}\sigma _{j}$, for which the mean field
approximation is exact and provides a critical temperature: $T_{c}=J/k_{B}.$%
The mapping is easily found by identifying $J/N$ with the effective NT-mediated H$_{2}-$H%
$_{2}$ interaction energy per particle: $v_{eff}=2[\epsilon -\epsilon
_{1}(d_{0})].$ For $^4$He, Ne and H$_2$ we find respectively $T_{c}=5.4$ K, $63.8$ K and $398.4$ K. 

To summarize, we have proposed that lattice dilation plays an important role
in gas uptake. Because the stimulus to this dilation is the resulting
increase in absorbate binding energy, the expansion is very sensitive to the
gas-tube interactions.The inter-tube and intermolecular interactions also
play an important role in determining the uptake threshold condition. We
have used plausible, but uncertain, semiempirical models for these
interactions. Experimental investigation (thermodynamic, diffraction and
Raman scattering) will shed light on these assumptions. The existence and nature of this dilation phenomenon and accompanying
transition are robust conclusions, evidently not dependent on the
approximations used in the interactions.

We are grateful to Vincent Crespi, Peter Eklund, Karl Johnson, Milen Kostov,
Ari Mizel, Paul Sokol and Keith Williams for stimulating conversations and
to Aldo Migone and Michel Bienfait for their communication of results prior
to publication. This research has been supported by the Petroleum Research
Fund of the American Chemical Society, the Army Research Office, Fundacion
Antorchas and CONICET.


\begin{references}
\bibitem{pressure}  This pressure may be estimated from either the 1D fluid
pressure, times the 2D IC density, or from the radial derivative of the
transverse eigenvalue, per unit confining area per particle
\end{references}
\end{document}